\newcommand{\npb}[3]{{\sl Nuc.~Phys.}\ {\bf B#1}, #2\ (#3)}
\newcommand{\prd}[3]{{\sl Phys.~Rev.}\ {\bf D#1}, #2\ (#3)}
\newcommand{\prl}[3]{{\sl Phys.~Rev.~Lett.}\ {\bf #1}, #2\ (#3)}
\newcommand{\plb}[3]{{\sl Phys.~Lett.}\ {\bf B#1}, #2\ (#3)}
\newcommand{\ijmpa}[3]{{\sl Int.~J.~Mod.~Phys.}\ {\bf A#1}, #2\ (#3)}

\newcommand{\jjlv}    {\mbox{$ j j \ell \nu                                $}}

\newcommand{\jjjj}    {\mbox{$ j j j j                                     $}}
\newcommand{\lvlv}    {\mbox{$ \ell \nu \ell \nu                           $}}
\newcommand{\ctw}     {\mbox{$ \cos\theta                                  $}}

\newcommand{\phia}    {\mbox{$ \phi_1                                      $}}
\newcommand{\cta}     {\mbox{$ \cos\theta_1                                $}}

\newcommand{\phib}    {\mbox{$ \phi_2                                      $}}
\newcommand{\ctb}     {\mbox{$ \cos\theta_2                                $}}
\newcommand{\ctl}     {\mbox{$ \cos\theta_l                                $}}
\newcommand{\phil}    {\mbox{$ \phi_l                                      $}}
\newcommand{\ctj}     {\mbox{$ \cos\theta_j                                $}}
\newcommand{\phij}    {\mbox{$ \phi_j                                      $}}
\newcommand{\ctja}     {\mbox{$ \cos\theta_{j_1}                           $}}
\newcommand{\phija}    {\mbox{$ \phi_{j_1}                                 $}}
\newcommand{\ctjb}     {\mbox{$ \cos\theta_{j_2}                           $}}
\newcommand{\phijb}    {\mbox{$ \phi_{j_2}                                 $}}
\newcommand{\fold}     {\mbox{$ _{\mathrm{folded}}                          $}}

\def\etal{{\it et al.}}
\def\vs{{\it vs}}
\def\eg{{\it e.g.}}
\def\etc{{\it etc.}}
\def\pbin{pb$^{-1}$}
\def\fbin{fb$^{-1}$}
\def\tchan{$t$-channel}

\def\sqrts{${\rm E}_{\rm CM}$}
\def\wm{W^-}
\def\wp{W^+}
\def\thew{\Theta_W}
\def\thed{\theta}
\def\phid{\phi}
\def\thedb{\bar\thed}
\def\phidb{\bar\phid}

\def\thetaw{\theta_W}
\def\sw{s_W}
\def\cw{c_W}
\def\cwsq{\cos^2\theta_W}
\def\swsq{\sin^2\theta_W}

\def\twsq{\tan^2\theta_W}

\def\lninel{L_{9L}}
\def\lniner{L_{9R}}

\def\partmud{\partial_\mu}

\def\Dmud{D_\mu}

\def\Dnud{D_\nu}

\def\mw{M_W}
\def\wwv{$WWV$}

\def\epem{$e^+e^-$}
\def\emem{$e^-e^-$}
\def\emg{$e^-\gamma$}
\def\gg{$\gamma\gamma$}
\def\epemww{$e^+e^-\rightarrow W^+W^-$}
\def\gonev{g_1^V}
\def\goneg{g_1^\gamma}
\def\gonez{g_1^Z}
\def\gfourv{g_4^V}
\def\gfivev{g_5^V}
\def\dgonez{\Delta\gonez}
\def\hiv{h_i^V}
\def\honev{h_1^V}
\def\htwov{h_2^V}
\def\hthreev{h_3^V}
\def\hfourv{h_4^V}
\def\kapv{\kappa_V}
\def\kapg{\kappa_\gamma}
\def\kapz{\kappa_Z}
\def\dkapv{\Delta\kapv}
\def\dkapg{\Delta\kapg}
\def\dkapz{\Delta\kapz}
\def\kapvtwid{\tilde{\kappa}_V}

\def\lamv{\lambda_V}
\def\lamg{\lambda_\gamma}
\def\lamz{\lambda_Z}
\def\lamvtwid{\tilde{\lambda}_V}

\def\doubarrow#1{#1\hskip-0.5em^{^\leftrightarrow}}

\def\alphawphi{\alpha_{W\phi}}
\def\alphabphi{\alpha_{B\phi}}
\def\alphaw{\alpha_{W}}

\def\eemr{e^-e^-\rightarrow}
\def\egr{e^-\gamma\rightarrow}
\def\ggr{\gamma\gamma\rightarrow}

\documentclass[12pt]{article}
\usepackage{graphics}

\begin{document}

\thispagestyle{empty}
\renewcommand{\thefootnote}{\fnsymbol{footnote}}

\begin{flushright}
{\small
hep-ph/9611454\\
MADPH-96-975 \\
SLAC-PUB-7366\\
UB-HET-96-05 \\
UM-HE-96-26 \\
November 1996\\}
\end{flushright}

\vfill

\begin{center}
{\bf\large   
Anomalous Gauge Boson Couplings\footnote{Work supported by
Department of Energy contract  DE--AC03--76SF00515.}}

\medskip

\def\barklow{a}
\def\baur{b}
\def\cuypers{c}
\def\dawson{d}
\def\erreded{e}
\def\erredes{\erreded}
\def\godfrey{f}
\def\han{g}
\def\kalyniak{\godfrey}
\def\riles{h}
\def\rizzo{\barklow}
\def\sobey{\han}
\def\strom{i}
\def\szalapski{j}
\def\ward{k}
\def\womersley{l}
\def\wudka{m}
\def\zeppenfeld{n}
T.~Barklow$^\barklow$, U.~Baur$^\baur$, F.~Cuypers$^\cuypers$, 
S.~Dawson$^\dawson$, D.~Errede$^{\erreded}$, \\
S.~Errede$^{\erredes}$\footnote{Subgroup conveners}, 
S.~Godfrey$^\godfrey$, T.~Han$^\han$, P.~Kalyniak$^\kalyniak$,
K.~Riles$^{\riles\dagger}$, \\
T.~Rizzo$^\barklow$, 
R.~Sobey$^\sobey$,
D.~Strom$^\strom$, R.~Szalapski$^\szalapski$, B.F.L.~Ward$^\ward$, \\
J.~Womersley$^\womersley$, 
J.~Wudka$^\wudka$, D.~Zeppenfeld$^\zeppenfeld$ \\ 
\medskip
{\it $^\barklow$Stanford Linear Accelerator Center, 
Stanford University, Stanford, CA 94309} \\
{\it $^\baur$State University of New York at Buffalo, 
Buffalo, NY 14260} \\
{\it $^\cuypers$Paul Scherrer Institute, 
CH-5232 Villigen PSI, Switzerland} \\
{\it $^\dawson$Brookhaven National Laboratory,
P.O. Box 5000, Upton, NY 11973} \\
{\it $^\erreded$University of Illinois at Urbana-Champaign, 
Urbana-Champaign, IL 61801} \\
{\it $^\godfrey$Carleton University,
Ottawa, ON K1S-5B6, Canada} \\
{\it $^\han$University of California at Davis,
Davis, CA 95616} \\
{\it $^\riles$University of Michigan, 
Ann Arbor, MI 48109} \\
{\it $^\strom$University of Oregon,
Eugene, OR 97403} \\
{\it $^\szalapski$KEK National Laboratory,
Tsukuba, Ibaraki 305, Japan} \\
{\it $^\ward$University of Tennessee,
Knoxville, TN 37996} \\
{\it $^\womersley$Fermi National Accelerator Laboratory,
Batavia, IL 60510} \\
{\it $^\wudka$University of California at Riverside,
Riverside, CA 92521} \\
{\it $^\zeppenfeld$University of Wisconsin,
Madison, WI 53706} \\

\end{center}

\vfill

\begin{center}
{\bf\large   
Abstract }
\end{center}

\begin{quote}
The measurement of anomalous gauge boson self couplings is
reviewed for a variety of present and planned
accelerators. Sensitivities are compared for these accelerators
using models based on the effective Lagrangian approach.
The sensitivities described here are for measurement of
``generic'' parameters $\kapv$, $\lamv$, \etc, defined in the text.
Pre-LHC measurements will not probe these coupling 
parameters to precision
better than $O(10^{-1})$. The LHC should be sensitive to better
than $O(10^{-2})$, while a future NLC should achieve sensitivity
of $O(10^{-3})$ to $O(10^{-4})$ for center of mass energies ranging from
0.5 to 1.5 TeV.
\end{quote}

\vfill

\begin{center} 
{\it Summary of the Snowmass Subgroup on Anomalous Gauge Boson Couplings, to appear in the} 
{\it Proceedings of the 
1996 DPF/DPB Summer Study on New Directions in
High-Energy Physics, June 25 - July 12, 1996, Snowmass,
CO, USA} \\
\end{center}

\newpage



%
\pagestyle{plain}

\section{Introduction}
Although the Standard Electroweak Model has been verified to astounding
precision in recent years at LEP and SLC, one important component
has not been tested directly with precision: the 
non-abelian self couplings of the weak vector gauge bosons. 
Deviations of non-abelian couplings from expectation would signal
new physics, perhaps arising from unexpected 
internal structure or loop corrections
involving propagators of new particles. In addition, as is
discussed elsewhere in these proceedings\cite{STRONG},
precise measurements of gauge boson self interactions
provide important information
on the nature of electroweak symmetry breaking.

Recent results from CDF and D0 are consistent with 
non-vanishing values of triple gauge boson couplings,
but have not yet reached a precision better than order unity.
Upcoming measurements at LEP II and later at an upgraded 
Tevatron will improve upon this precision by an order of magnitude.
The Large Hadron Collider (LHC) should do better by more 
than another order of magnitude. A 500 GeV
Next Linear Collider (NLC) improves still further upon the
LHC precision, and a 1.5 TeV NLC probes even smaller couplings,
of order $10^{-4}$. The alternative colliding modes of an
NLC machine, $e^-e^-$, $e^-\gamma$, and $\gamma\gamma$, also provide
useful and often complementary information on anomalous
couplings. In principle, 
a $\mu^+\mu^-$ collider would provide 
comparable sensitivity to a corresponding NLC $e^+e^-$ 
machine of the same 
center of mass (c.m.) 
energy, as long as luminosities, beam polarization, and
detector backgrounds are also kept comparable, requirements
that seem daunting at the moment. It should be noted, though,
that given the increased sensitivity to anomalous couplings that
comes with higher c.m. energies, a 4 TeV $\mu^+\mu^-$ collider
would be a powerful machine indeed for studying gauge boson
self-interactions.

This article focuses on {\it direct} measurements of anomalous
couplings, typically via diboson production. There also exist
indirect, model-dependent limits,
obtained from virtual corrections to precisely
measured observables and inferred parameters, such as 
$(g-2)_\mu$, the neutron electric dipole moment,
and ``oblique'' $Z$ parameters. 
Depending on assumptions,
\eg, depending on what one considers ``natural'',
one can obtain limits from $O(10^{-2})$ to 
$O(1)$ \cite{HINCHCLIFFE}.

In the following sections, an overview is given of common 
parametrizations of anomalous
gauge boson couplings, followed by discussions of sensitivities
to anomalous parameters provided by various accelerators.
Conclusions follow at the end.

\section{Parametrization}
Anomalous gauge boson couplings can be parametrized in a variety
of ways. One can define ``generic'' parameters that describe
in the most general way the allowed forms of gauge-boson
vertices. This generic form has many free parameters,
some of which violate discrete symmetries.
To deal with this multitude
of parameters it is convenient to apply an effective Lagrangian
approach, assume certain symmetries are respected, and expand
in terms of given operator dimensions. This approach has the virtue
of imposing relations among the many otherwise-arbitrary parameters,
and allowing an {\it a priori} estimate of the relative 
importance of different contributions.

In defining a generic set of anomalous couplings, 
we follow the notation of
ref.~\cite{HHPZ} in which the effective Lagrangian for
the \wwv\ vertex is written:
\begin{eqnarray*}
L_{WWV} / g_{WWV} 
& = &     i\,\gonev(W^\dagger_{\mu\nu}W^\mu V^\nu 
                  - W^\dagger_\mu V_\nu W^{\mu\nu}) \\
& &      + i\,\kapv W^\dagger_\mu W_\nu V^{\mu\nu} \\
& &      + {i\,\lamv\over\mw^2}W^\dagger_{\lambda\mu}
         W^\mu_{\>\nu}V^{\nu\lambda} \\
& &   - \gfourv W^\dagger_\mu W_\nu(\partial^\mu V^\nu
                                    +\partial^\nu V^\mu) \\
& &      + \gfivev\epsilon^{\mu\nu\rho\sigma}
        (W^\dagger_\mu\doubarrow{\partial}_\rho W_\nu)
        V_\sigma \\
& &      + \kapvtwid W^\dagger_\mu W_\nu\tilde{V}^{\mu\nu}\\
& &   + {i\,\lamvtwid\over\mw^2}W^\dagger_{\lambda\mu}
        W^\mu_{\>\nu}\tilde{V}^{\nu\lambda} 
\end{eqnarray*}
where $W_{\mu\nu}\equiv\partial_\mu W_\nu -\partial_\nu W_\mu$,
$V_{\mu\nu}\equiv\partial_\mu V_\nu -\partial_\nu V_\mu$,
$(A\doubarrow{\partial_\mu}B)\equiv 
A(\partial_\mu B)-(\partial_\mu A)B$, and
$\tilde{V}_{\mu\nu} \equiv 
{1\over2}\epsilon_{\mu\nu\rho\sigma}V^{\rho\sigma}$. The
normalization factors are defined for convenience to be
$g_{WW\gamma} \equiv -e$ and $g_{WWZ} \equiv -e\cot\theta_W $.
The 14 (7$\times$2) general coupling parameters allow for
C/P-violating ($\gfivev)$, 
and CP-violating ($\gfourv,\kapvtwid,\lamvtwid$) terms. In most studies
such terms are neglected. In the standard model at tree level
$\gonev=\kapv=1$ and $\lamv=\gfourv=\gfivev=\kapvtwid=\lamvtwid=0$.
It should be noted that these couplings are, in general, form
factors with momentum-dependent values. This complication is of little 
importance at an \epem\ collider where the $WW$ c.m. energy is well
defined, but it must be borne in mind at hadron colliders where
couplings are simultaneously probed over 
large energy ranges\cite{ZEPPENFELD}. To observe the
momentum dependence directly at a hadron collider requires
a nominal c.m. energy comparable to the form factor scale parameter 
$\Lambda_{\rm FF}$.

We follow a common convention in defining 
$\dgonez \equiv \gonez - 1$ and $\dkapv \equiv \kapv - 1$.
The $W$ electric charge fixes $\goneg(q^2\rightarrow0) \equiv 1$.
In perhaps more familiar notation, 
in the static limit one can express the $W$
magnetic dipole moment as 
$$ \mu_W \quad \equiv\quad {e\over2\,\mw}(1+\kapg+\lamg)$$
\noindent and the $W$ electric quadrupole moment as
$$ Q_W \quad\equiv\quad-{e\over\mw^2}(\kapg-\lamg).$$

In addition, one can investigate the tri-boson coupling at the
$V_\mu(P)\rightarrow Z_\alpha(q_1)\gamma_\beta(q_2)$ vertex
(with $V=\gamma,Z$) for which the following general vertex function 
can be written:
\begin{eqnarray*}
\Gamma^{\alpha\beta\mu}_{Z\gamma V} \quad=\quad & {s-M_V^2\over M^2_Z}
\bigl[\>&\honev(q_2^\mu g^{\alpha\beta}-q_2^\alpha g^{\mu\beta}) + \\
& & {\htwov\over M_Z^2}P^\alpha(P\cdot q_2g^{\mu\beta}-
q_2^\mu P^\beta) +\\
& & \hthreev\epsilon^{\mu\alpha\beta\rho}q_{2_\rho} +\\
& & {\hfourv\over M_Z^2}P^\alpha\epsilon^{\mu\beta\rho\sigma} P_\rho
q_{2_\sigma}\>\bigr]
\end{eqnarray*}
where $\hiv$ are anomalous couplings (zero in the Standard Model
at tree level).
The couplings $\honev$ and $\htwov$ are CP-violating and are
typically ignored in studies.
As is true for the generic $WWV$ coupling parameters, these
parameters are, in general, momentum-dependent 
form factors.

There are two common approaches taken in relating these generic
parameters to those of effective Lagrangians
that go beyond the Standard Model.
Both approaches involve classifying Lagrangian terms 
according to the energy dimensions of the operators involved,
where each term beyond dimension 4 is suppressed by
a power of a large mass parameter $\Lambda$ that characterizes
the scale of new physics: $L_{\rm eff} = L_{\rm SM} + L_{\rm NR}$
where $L_{\rm NR}$ is non-renormalizable in finite 
order:\cite{EINHORNHAWAII,WUDKA}
\begin{eqnarray*}
L_{NR} & \equiv & 
{1\over\Lambda}\sum_i \alpha_i^{(5)} O_i^{(5)} +  \\
& & {1\over\Lambda^2}\sum_i \alpha_i^{(6)} O_i^{(6)} + ... 
\end{eqnarray*}
where $O_i^{N}$ are local operators of dimension $N$ and
$\alpha_i^{(N)}$ are dimensionless couplings. 
Since odd-dimension operators do not contribute to gauge boson
self interactions, one begins with dimension-6
operators and typically assumes that dimension-8 operators
can be neglected. 

In the so-called linear realization,
(in which the Standard Model is recovered in the limit
$\Lambda\rightarrow\infty$), one includes in the Lagrangian 
an explicit Higgs doublet field and its associated covariant
derivative. Terms are then separated into those affecting
gauge-boson two-point functions, which have been well tested
at LEP/SLC, and those leading to non-standard triple
gauge boson couplings. One generally expects in this model 
suppressions of dimension-6 anomalous
triple gauge terms by $O(\mw^2/\Lambda^2)$.
It has been shown\cite{AEW}, however, that no renormalizable
underlying theory can generate these terms at tree level. One
requires loops and thereby incurs an additional suppression of
$O({1\over16\,\pi^2})$. In this scheme it is difficult to
observe large anomalous trilinear couplings without
directly observing the new physics itself. 
In contrast, anomalous quartic couplings can be generated at
tree level and may therefore be substantially larger than
the anomalous trilinear couplings. In the most
widely used linear realization, one assumes (somewhat arbitrarily)
the additional constraint of equal couplings for the
$U(1)$ and $SU(2)$ terms in the 
effective Lagrangian that contribute to
anomalous triple gauge boson couplings. 
One also neglects C, P, and CP-violating terms in the Lagrangian.
This leads to the
``HISZ Scenario'', named after the authors\cite{HISZ}, and involves
only two free parameters, commonly taken to be $\kapg$ and
$\lamg$. In this scenario, the following relations hold:
\begin{eqnarray*}
\dkapz & = & {1\over2}(1-\tan^2\thetaw)\dkapg \\
\dgonez & = & {1\over2\,\cos^2\thetaw}\dkapg \\
\lamz & = & \lamg 
\end{eqnarray*}
For reference, a ``relaxed'' HISZ scenario is sometimes
used (see LEP II studies below) in which the $U(1)$ and $SU(2)$
couplings are {\it not} equated, a scenario that then involves
three free parameters.

In the second common approach to effective Lagrangians,
one constructs a nonlinear field from would-be Goldstone
Bosons which give masses to the 
$W$ and $Z$ Bosons\cite{BAGGER}. Without an
explicit Higgs doublet in the model, one must
encounter new physics at the few-TeV scale in a process
such as longitudinal $W$-$W$ scattering, which would
otherwise violate unitarity. This non-linear approach is
discussed in detail elsewhere in these proceedings\cite{STRONG}.
For completeness, salient features are outlined here.

A non-linear field is defined via
$$ \Sigma \quad \equiv \quad e^{i\vec w\cdot\vec\sigma/v}$$
with covariant derivative:
$$ \Dmud\Sigma \quad = \quad \partmud\Sigma + {i\over2}g
W^a_\mu\sigma^a\Sigma-{i\over2}g'B_\mu\Sigma\sigma_3 $$
where $w_i$ represent the Goldstone fields, $\sigma_i$ are the
Pauli spin matrices, and
$v=246$ GeV is the electroweak symmetry breaking scale. 
From these fields and covariant derivatives, an effective
Lagrangian is constructed, for which 
the following dimension-6 terms give anomalous $WWV$ couplings:
\begin{eqnarray*}
&& -ig{v^2\over\Lambda^2}\lninel Tr[W^{\mu\nu}\Dmud\Sigma
\Dnud\Sigma^\dagger]\quad + \\
&& -ig'{v^2\over\Lambda^2}\lniner Tr[B^{\mu\nu}\Dmud\Sigma^\dagger
\Dnud\Sigma]
\end{eqnarray*}
In this scheme, the coupling parameters $\lninel$ and $\lniner$
are expected to be $O(1)$. They can be mapped onto the
generic set of parameters defined above via the relations:
\begin{eqnarray*}
\dgonez \quad & = & \quad {e^2\over2\,\cw^2\sw^2} {v^2\over\Lambda^2}
\>\lninel \\
\noalign{\bigskip}
\dkapg \quad & = & \quad  {e^2\over2\,\sw^2} {v^2\over\Lambda^2}
\>(\lninel+\lniner) \\
\noalign{\bigskip}
\dkapz \quad & = & \quad  {e^2\over2\,\cw^2\sw^2} {v^2\over\Lambda^2}
\>(\lninel\,\cw^2-\lniner\,\sw^2) 
\end{eqnarray*}
where $\sw^2=\swsq$ and $\cw^2=\cwsq$.
In the non-linear scheme there are no $\lamv$ terms at the 
dimension 6 level.
It should be noted that one obtains the full HISZ scenario with 
$\lamg=0$ by setting $\lniner=\lninel$. 

\section{Present Limits on Anomalous Gauge Couplings}
\begin{figure*}[b]
\begin{center}
\leavevmode
\resizebox{3.5in}{!}
{\includegraphics{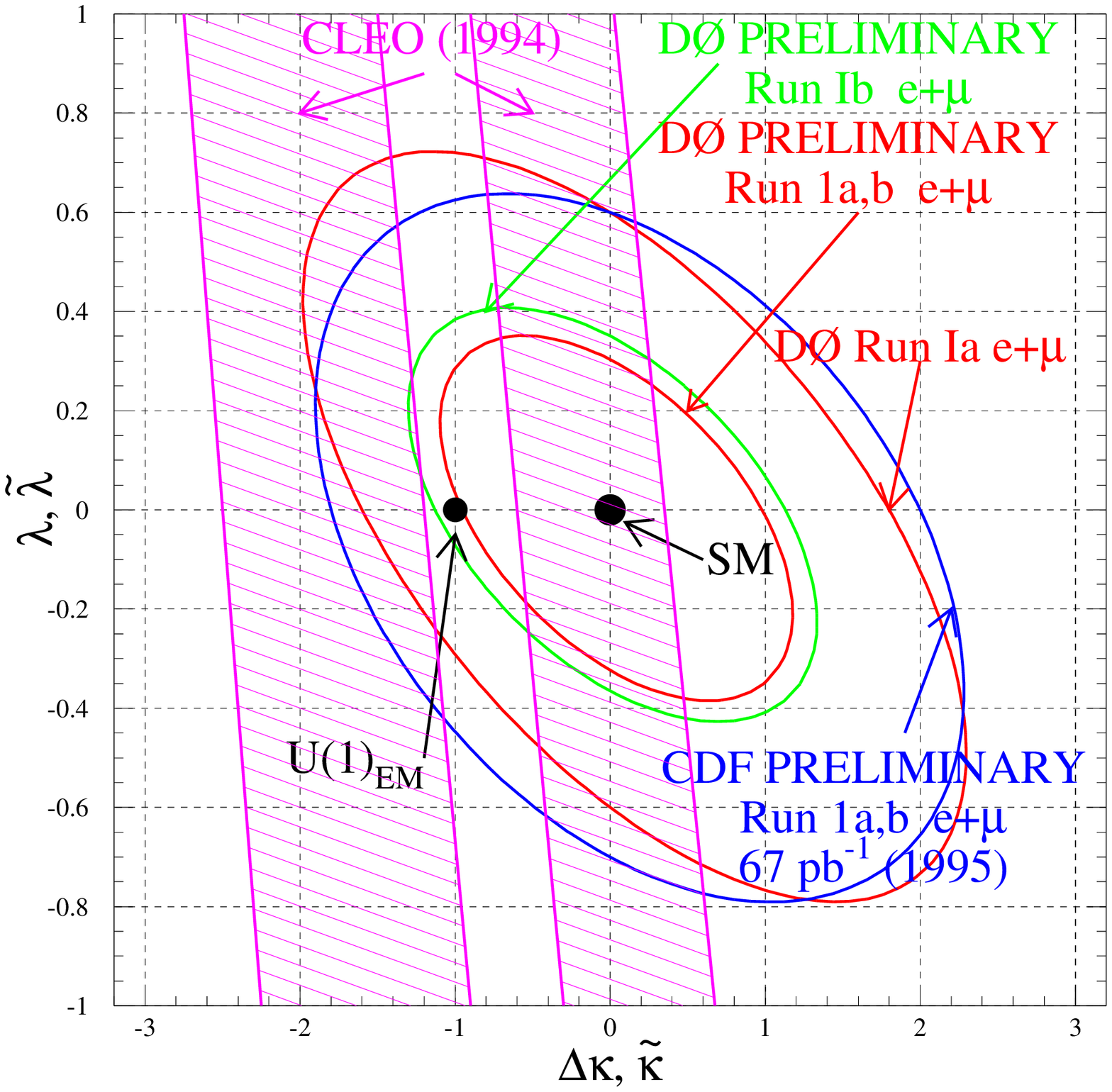}}
\end{center}
\caption{CDF and D0 95\% C.L. contours for \wwv\ anomalous couplings
for $\Lambda$ = 1.5 TeV. Also shown are the CLEO limits from
$b \rightarrow s \gamma/B \rightarrow K^* \gamma$.}
\label{Wgam_limits}
\end{figure*}

The best present direct measurements of \wwv\ couplings come
from hadron collider experiments $-$ the
UA2 experiment at the CERN S$\bar{p}p$S\cite{UA2} and 
the CDF and D0 experiments at the Fermilab Tevatron\cite{Wgamma,WW_WZ,AGBI}.
UA2 has searched for $W\gamma$ production, while CDF and D0 have searched for
$W\gamma$, $WW$ and $WZ$ production. 
Preliminary CDF and D0 results with $\sim$ 100~pb$^{-1}$ of 
Run 1 data have
resulted in $O(5)$ $WW$ candidates per experiment in the dilepton channel
and $O(100)$ $W\gamma$ candidates per experiment in the $e$ and $\mu$ channels,
consistent with SM expectations. From these data, the absence of an excess of
such events has resulted in 95\% C.L. bounds 
on coupling parameters of the 
order of unity, as shown in Fig.~\ref{Wgam_limits}. 
(Also shown are limits from the CLEO experiment determined from
measurements of B($b\rightarrow s\gamma$)\cite{CLEO}.)
For example, from $W\gamma$ data, D0 sets 95\% C.L. limits of 
$|\dkapg| < 1.0$ assuming $\lamg=0$ and 
$|\lamg|  < 0.3$ assuming $\dkapg=0$ (both results 
assume $\Lambda=1.5$ TeV 
in the appropriate form factors). In the static limit,
these constraints on $WW\gamma$ 
anomalous couplings can be related to 
higher-order EM moments of the $W$ boson - the magnetic dipole moment
and electric quadrupole moment, both of which are predicted to be non-zero
in the SM. These experimental results exclude the simultaneous vanishing of
$\mu_W$ and $Q_W$ in excess of 95\% C.L..

Preliminary results from CDF and D0 on limits on \wwv\ anomalous couplings from 
$WW \rightarrow \ell \nu jj$ production from Run 1 data are comparable to 
those obtained from $W\gamma$ production. Limits from the $WW$ dilepton channel
are approximately 60\% higher from each experiment. Since
$WW/WZ$ production is sensitive to both $WW\gamma$ and $WWZ$ couplings, and
delicate gauge cancellation between the two is required by the SM, these 
results are interesting because they provide the first {\it direct}
evidence of the existence of the $WWZ$ coupling $-$ i.e. 
$g_1^Z = \kappa^Z = 0$ is excluded in excess of 99\% C.L..

\begin{figure*}[hbt]
\begin{center}
\leavevmode
\resizebox{3.5in}{!}
{\includegraphics{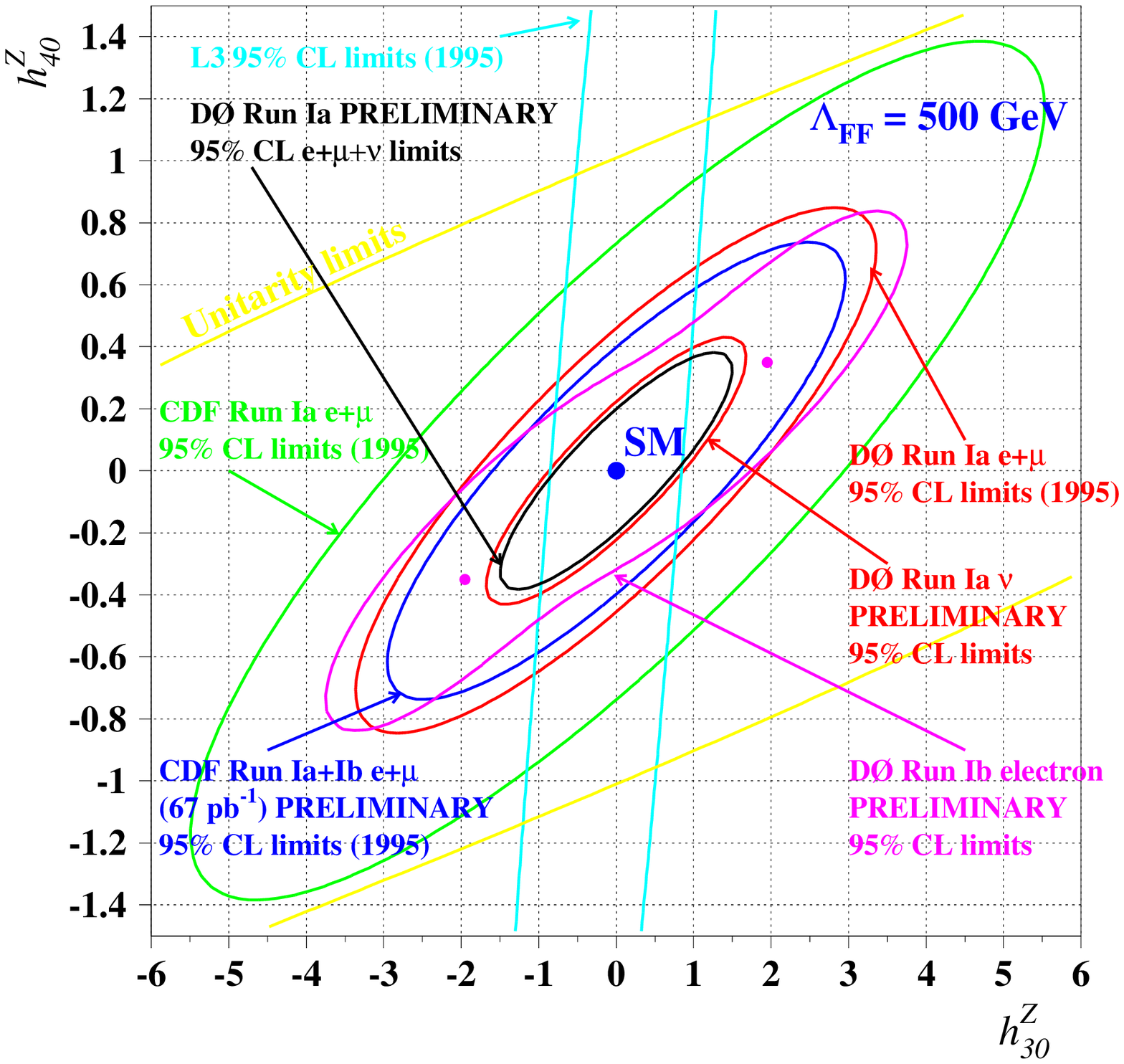}}
\end{center}
\caption{CDF and D0 95\% C.L. contours for $Z\gamma V$ anomalous couplings
for $\Lambda$ = 0.5 TeV}
\label{Zgam_limits}
\end{figure*}

CDF, D0 and the L3 and DELPHI experiments 
at LEP have also obtained direct limits on 
$Z\gamma V$ anomalous couplings\cite{Zgamma,L3,DELPHI}. 
The SM predicts these
couplings to be zero at tree level. At the Tevatron, the 
$Z\gamma \rightarrow \ell\ell\gamma$ ($\ell = e, \mu$) 
channel has been used, and more than $O$(10) of $Z\gamma$ 
candidates have been observed in Run 1 data. 
The L3 experiment has searched in the lepton channel as well as the 
$\nu {\bar \nu} \gamma$ channel. 
D0 has also searched in this latter channel, 
and extracted the most stringent preliminary 95\% C.L. limits of 
$|h^V_{30,10}| < 0.9$ for $h^V_{40,20} = 0$ and
$|h^V_{40,20}| < 0.2$ for $h^V_{30,10} = 0$, for $\Lambda=0.5$ TeV. 
The limits from the $\ell \ell \gamma$ channels 
from each experiment are 
approximately a factor of two less 
stringent than that obtained from the 
$\nu {\bar \nu} \gamma$ channel D0 result, as shown in Fig.~\ref{Zgam_limits}. 

\section{What Can LEP II Say?}

One expects significant improvement at 
LEP II\cite{LEPII,GODFREYLEP2}, as
the accelerator raises its energy to well above the $W$-pair 
production threshold. Results from the recent
Physics at LEP II Workshop\cite{LEPII} will be summarized here.
In the workshop study, the relaxed
HISZ scenario was used.
Three free parameters $\alphawphi$,
$\alphabphi$, and $\alphaw$ were considered, which can be
related to the generic set via the relations:
\begin{eqnarray*}
\dkapg & = & 
\alphawphi + \alphabphi \\
\dkapz & = & \alphawphi - \twsq\alphabphi \\
\dgonez  & = &  {1\over\cwsq}\>\alphawphi \\
\lamg & = & \lamz\quad =\quad \alphaw
\end{eqnarray*}
(If one requires $\alphawphi$ = $\alphabphi$, one recovers the
full HISZ scenario.)

As for hadron colliders, lepton colliders provide sensitivity to
anomalous couplings through studies of diboson production.
One important advantage of lepton colliders, though, is the
ability to reconstruct accurately the full kinematics of the
$W$ pair events because of the absence of spectator partons. 
To a good approximation,
full energy and momentum conservation constraints can be applied to the
visible final states. 
Thus an \epemww\ event can ideally be characterized by five angles:
the production angle $\thew$ of the $\wm$ with respect to the electron beam,
the polar and azimuthal decay 
angles $\thed$ and $\phid$ of one daughter
of the $\wm$ in the $\wm$ reference frame, and the corresponding
decay angles $\thedb$ and $\phidb$ of a $\wp$ daughter. 
In practice, initial-state photon radiation and final-state
photon and gluon radiation (in hadronic W decays) complicate the
picture. So does the finite width of the $W$. 
Nevertheless, one finds that the five reconstructed 
angles remain robust observables for studying anomalous couplings.

There are three main topologies to consider, those in which
both $W$'s decay hadronically, in which 
one decays hadronically, the other 
leptonically, and in which both decay leptonically. 
The fully hadronic topology has the advantages of abundance 
(45.6\%) and fully measured kinematics, but has the disadvantage
of poorly determined charge signs for the $W$'s and especially
their decay products. Since much of the sensitivity to anomalous
couplings comes from the pronounced forward-backward charge
asymmetry of the $W$'s (distribution in $\thew$), this disadvantage
is a serious one. The mixed hadronic/leptonic topology has the
disadvantages of lower abundance (29.2\%, counting only $e$, $\mu$
decays) and of more poorly measured kinematics, due to the missing
neutrino. This latter disadvantage is largely offset, however, by
the power of kinematic constraints, including energy/momentum
constraints and (optionally) mass constraints on reconstructed
$W$'s. The considerable advantage of the mixed topology lies in
unambiguous determination of the individual $W$ electric charges
and the charge of one decay product. There does remain, however,
a two-fold ambiguity in decay angles of the hadronic final state.
The purely leptonic topology suffers badly from its low
abundance (10.5\%, counting $e$, $\mu$ and $\tau$ decays) and
from the very poorly known kinematics, given two missing neutrinos.
Even after imposing energy/momentum constraints and forcing
both $W$ masses to reconstruct to the nominal value, one is still
left with a two-fold ambiguity in resulting angles.

In the LEP II study,
a number of fitting methods were considered in order to extract
anomalous couplings from measured angular distributions,
methods based on helicity-density matrices, maximum likelihood,
and moments of multi-dimensional distributions. The maximum 
likelihood technique was found to be most effective, and results
from that method are shown here. For the purely hadronic and the
mixed topologies, more than one choice of measured distributions 
for fitting was considered. 
Table~\ref{TABLEPII}~\cite{LEPII,SEKULIN} shows 
expected 1 standard deviation 
errors on each $\alpha$ parameter when only that
parameter is allowed to vary. The numbers represent integrated
luminosities of 500 \pbin\ at 176 and 190 GeV, respectively. For
comparison, the ``ideal'' sensitivity is also shown, for which
all five angles are reconstructed perfectly with no ambiguity.
From this table it's clear that the mixed topology, using
all five reconstructed angles provides the best sensitivity.
It's also clear that even a modest increase in energy improves
sensitivity significantly. From these numbers, one can expect
ultimate sensitivities (95\% C.L.) to anomalous couplings parameters
at somewhat better than $O(10\%)$.
It should be noted that when more than one coupling parameter
is varied at a time, correlations degrade these limits
significantly. The absence of beam polarization at LEP makes
separation of $WW\gamma$ from $WWZ$ couplings more difficult
than should be possible at the NLC, as discussed below.
\begin{table*}[h]
\begin{center}
\caption{Estimated 1 s.d. errors on anomalous couplings obtainable
at LEP II with 500 \pbin\ at each c.m. energy.}
\label{TABLEPII}
\begin{tabular}{|c|c|l|c|c|}
\hline\hline
Model & Channel   &    Angular data used  & 176 GeV & 190  GeV \\ \hline\hline
$\alpha_{B\phi}$
      & \jjlv     &      \ctw             & 0.222    & 0.109   \\
      &           &  \ctw, (\ctl, \phil)  & 0.182    & 0.082   \\
      &           &  \ctw, (\ctl, \phil),
                     (\ctj, \phij)\fold   & 0.159    & 0.080   \\
      & \jjjj     &  $\mid\ctw\mid$       & 0.376    & 0.149   \\
      &           &  $\mid\ctw\mid$,
                    (\ctja, \phija)\fold,
                   (\ctjb, \phijb)\fold   & 0.328    & 0.123   \\ 
      & \lvlv     &  \ctw, (\cta, \phia),
              (\ctb, \phib), 2 solutions  & 0.323    & 0.188   \\
      & Ideal     &  \ctw, (\cta, \phia),
                     (\ctb, \phib)        & 0.099    & 0.061   \\ \hline\hline
$\alpha_{W\phi}$
      & \jjlv     &   \ctw                & 0.041    & 0.027   \\
      &           &  \ctw, (\ctl, \phil)  & 0.037    & 0.023   \\
      &           &  \ctw, (\ctl, \phil), 
                     (\ctj, \phij)\fold   & 0.034    & 0.022   \\
      & \jjjj     &  $\mid\ctw\mid$       & 0.098    & 0.054   \\
      &           &  $\mid\ctw\mid$,
                    (\ctja, \phija)\fold,
                   (\ctjb, \phijb)\fold   & 0.069    & 0.042   \\ 
      & \lvlv     &  \ctw, (\cta, \phia),
              (\ctb, \phib), 2 solutions  & 0.096    & 0.064   \\
      & Ideal     &  \ctw, (\cta, \phia),
                     (\ctb, \phib)        & 0.028    & 0.018   \\ \hline\hline
$\alpha_W$
      & \jjlv     &      \ctw             & 0.074    & 0.046   \\
      &           &  \ctw, (\ctl, \phil)  & 0.062    & 0.038   \\
      &           &  \ctw, (\ctl, \phil),
                     (\ctj, \phij)\fold   & 0.055    & 0.032   \\
      & \jjjj     &  $\mid\ctw\mid$       & 0.188    & 0.110   \\
      &           &  $\mid\ctw\mid$,
                    (\ctja, \phija)\fold,
                   (\ctjb, \phijb)\fold   & 0.131    & 0.069   \\ 
      & \lvlv     &  \ctw, (\cta, \phia),
              (\ctb, \phib), 2 solutions  & 0.100    & 0.064   \\
      & Ideal     &  \ctw, (\cta, \phia),
                     (\ctb, \phib)        & 0.037    & 0.022   \\ \hline
\end{tabular}

\end{center}
\end{table*}

\section{What Can the Tevatron with the Main Injector Say?}
After the Main Injector upgrade has been completed, it is expected
that the Tevatron will collect $O(1$-$10)$ \fbin\ of data. (Further
upgrades in luminosity are also under discussion.)
If 10 \fbin\ is achieved, it is expected\cite{AGBI,Wendt} that
limits on $\dkapg$ and $\lamg$ will be obtained that are competitive
with those from LEP II with 500 \pbin\ at \sqrts\ = 190 GeV. 

In the Main Injector era, the Tevatron also provides a unique opportunity for
observing the SM prediction of the existence of a radiation amplitude zero in 
$W\gamma$ production\cite{RAZ}. Direct evidence for the existence of this
amplitude zero in $W\gamma$ production can be obtained by studying the
photon $-$ $W$-decay lepton rapidity correlation, or equivalently, the
photon-lepton rapidity difference distribution\cite{BEL}. 
Because of the fact that the LHC will be a $pp$ machine and 
because of severe QCD corrections at very high energies,
this will be an 
extremely difficult measurement at the LHC.

Limits on $Z\gamma V$ couplings for the same integrated luminosity, 
in the $\nu {\bar \nu} \gamma$ channel are anticipated to be 
$|h^V_{30,10}| < 0.024$ for $h^V_{40,20} = 0$ and
$|h^V_{40,20}| < 0.0013$ for $h^V_{30,10} = 0$, for $\Lambda=1.5$ TeV at
95\% C.L.. For the $\ell \ell \gamma$ channel, the limits on
$Z\gamma V$ anomalous couplings are expected to be
approximately a factor of two less stringent than this.

\section{What Can the LHC Say?}
One expects the LHC accelerator to turn on sometime
before the NLC and to look for the same signatures considered at
the Tevatron. The planned luminosity and c.m. energy, however,
give the LHC a large advantage over even the Main Injector Tevatron
in probing anomalous couplings. The ATLAS Collaboration has
estimated\cite{ATLAS} that with 100 \fbin, one can obtain 
(in the HISZ scenario) 95\% C.L. limits on $\dkapg$ and $\lamg$
in the range 5-10 $\times$ $10^{-3}$. It should be noted that
these studies do not yet include helicity analysis on the
$W$ bosons. A study of $WZ$ production at the LHC, in the all-lepton
decay channel\cite{AGBI,Womersley}, obtained similar results. For the $W\gamma$
channel, the limits on $\lamg$ are comparable, while the limits on
$\dkapg$ are approximately a factor of 10 times weaker.

The limits on $Z\gamma V$ couplings that are achievable at the LHC with
the same integrated luminosity are
$|h^V_{30,10}| < 5\times 10^{-3}$ for $h^V_{40,20} = 0$ and
$|h^V_{40,20}| < 9\times 10^{-4}$ for $h^V_{30,10} = 0$, 
for $\Lambda=1.5$ TeV\cite{AGBI}. It should be borne in mind that
these limits depend strongly upon the assumed form factor scale
$\Lambda$.

\section{What Can the NLC Say?}

A high-energy NLC will be able to replicate the measurements
of anomalous $WWV$ couplings expected at LEP II, but with two important
advantages: much higher energy and high beam polarization.
The increased energy allows dramatic improvement in sensitivity,
reflecting the fact that, in the effective 
Lagrangian description, the anomalous couplings arise from
higher-dimension effective interactions. The beam polarization
allows a clean separation of effects due to $WW\gamma$ and
$WWZ$ interactions. As at LEP II, one relies on reconstructing
(with additional help from kinematic constraints) the
five production / decay angles characterizing an
\epemww\ event. The resulting angular distributions are then
fitted to extract anomalous couplings.

At high energies in the Standard Model, 
the \epemww\ process is dominated by \tchan\
$\nu_e$ exchange, leading primarily to very forward-angle 
$W$'s where the $\wm$ has an average helicity near minus one.
This makes the bulk of the cross section difficult to observe
with precision. However, the amplitudes affected by the anomalous
couplings are not forward peaked; the central and backward-scattered
$W$'s are measurably altered in number and helicity by the couplings.
$W$ helicity analysis through the decay angular distributions provides
a powerful probe of anomalous contributions. 
Most detailed studies to date have restricted attention to 
events for which $|\cos\thew| <0.8$ and to the mixed topology
where one $W$ decays hadronically and the other leptonically.

\begin{figure*}[b]
\begin{center}
\leavevmode
\includegraphics{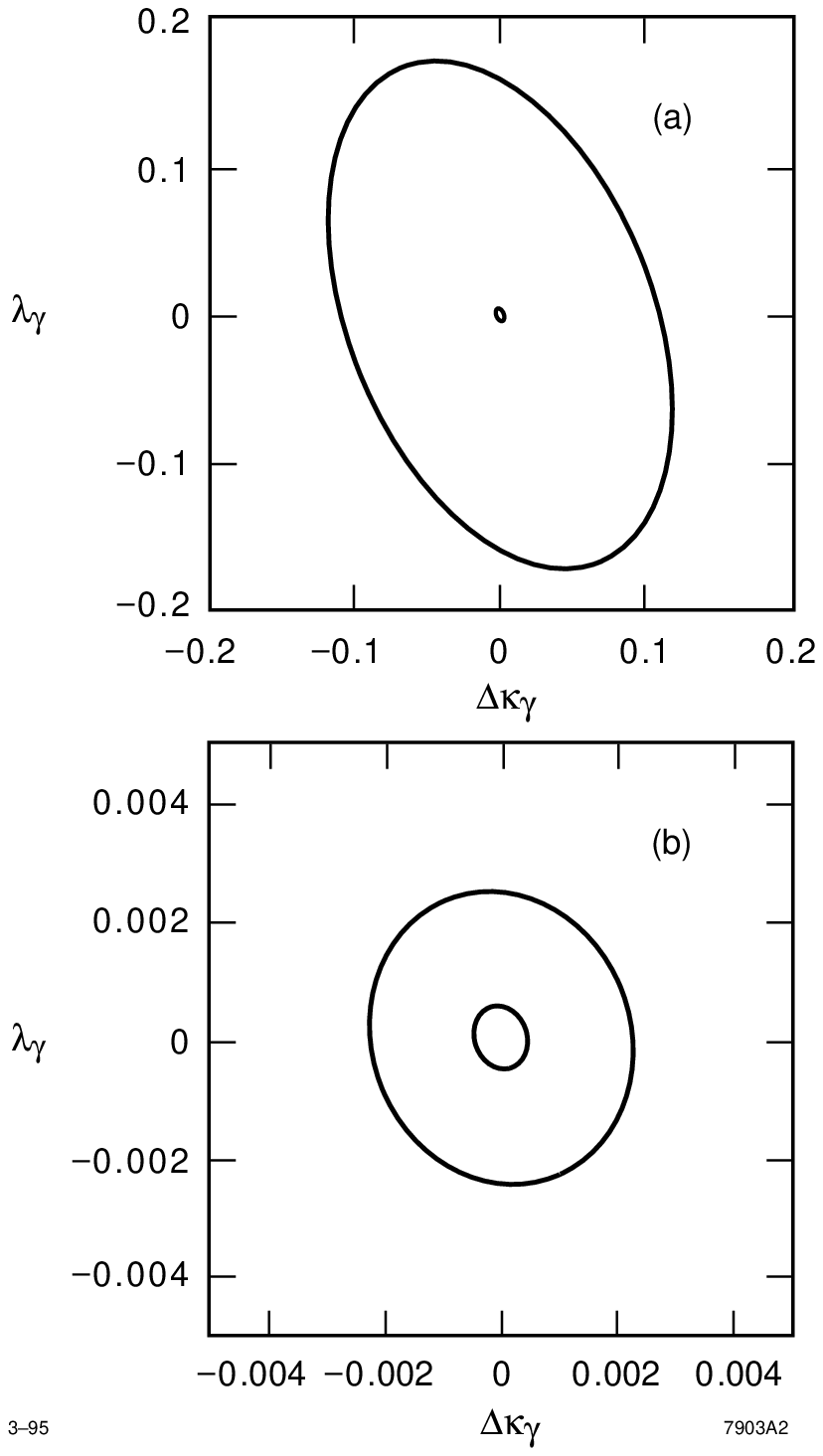}
\end{center}
\caption{95\% C.L. contours for $\dkapg$ and $\lamg$ in the
HISZ scenario. The outer contour in (a) is for \sqrts\ = 190 GeV
and 0.5 \fbin. The inner contour in (a) and the outer contour
in (b) is for \sqrts\ = 500 GeV with 80 \fbin. The inner
contour in (b) is for \sqrts\ = 1.5 TeV with 190 \fbin.}
\label{WWVFIGTWO}
\end{figure*}

\begin{figure*}[b]
\begin{center}
\leavevmode
\includegraphics{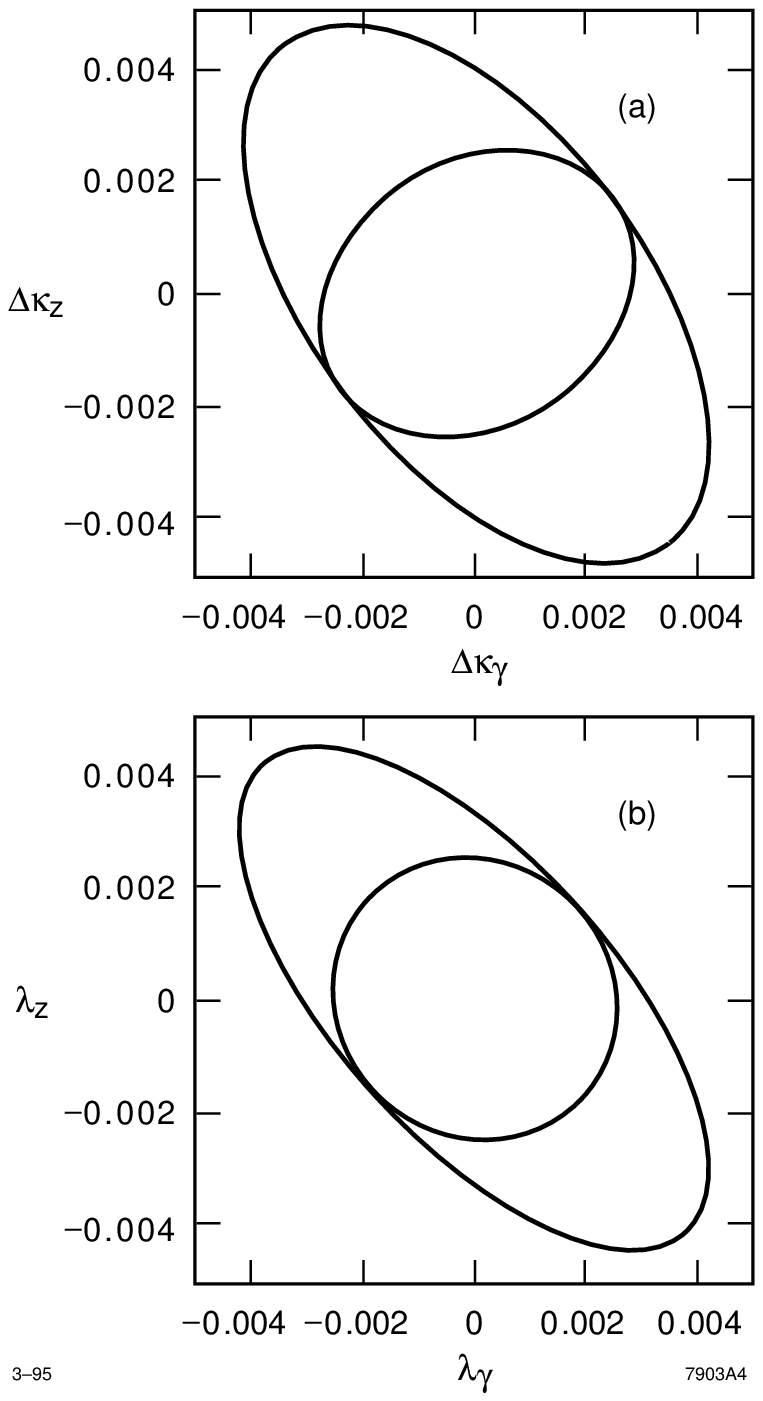}
\end{center}
\caption{95\% C.L. contours for $\dkapg$ and $\dkapz$
and for $\lamg$ and $\lamz$ for simultaneous fits of
$\dkapg,\dkapz,\lamg,\lamz$ at \sqrts\ = 500 GeV with
80 \fbin. The outer contours are for 0\% initial state
electron polarization and the inner contours are for
90\% polarization.}
\label{WWVPOLAR}
\end{figure*}

Fig.~\ref{WWVFIGTWO} (taken from ref.~\cite{TLBUCLA})
shows results from one such study. The figure depicts 
95\% C.L. exclusion contours in the plane
$\lamg$ \vs\ $\dkapg$ in the HISZ scenario for different c.m.
energies and integrated luminosities (0.5 \fbin\ at 190 GeV,
80 \fbin\ at 500 GeV, and 190 \fbin\ at 1500 GeV). These contours
are based on ideal reconstruction of $W$ daughter pairs
produced on mass-shell with no initial-state radiation (ISR). The
contours represent the best one can do. Another
study\cite{TLBFINLAND} assuming a very high-performance 
detector but including initial-state radiation and a finite
$W$ width found some
degradation in these contours, primarily due to efficiency
loss when imposing kinematic requirements to suppress events
far off mass-shell or at low effective c.m. energies.
Nevertheless, one attains a precision of $O(10^{-3})$ at NLC(500)
and $O($few $\times10^{-4})$ at NLC(1500). As mentioned above,
the polarizable beams at NLC allow one to
disentangle couplings that have correlated effects on observables
in accelerators with unpolarized beams. This feature facilitates
studying models more general than, say, the HISZ scenario with
only two free parameters. 
Fig.~\ref{WWVPOLAR} (taken from ref.~\cite{TLBUCLA})
shows expected 95\% C.L. exclusion contours in the
$\dkapg$--$\dkapz$ plane when $\dkapg$, $\lamg$, $\dkapz$,
and $\lamz$ are allowed to vary freely. The outer contour
is for unpolarized beams, while the inner contour where 
correlation is much reduced demonstrates the discrimination
available with 90\% beam polarization. Results consistent
with these have been found by other recent 
studies\cite{GODFREY3,HAN}. 

A more recent study \cite{KR} undertaken for 
the 1995-1996 NLC workshop\cite{NLCWORKSHOP} has examined
the effects of realistic detector resolution on achievable precisions.
One might expect {\it a priori} that the charged track momentum
resolution would be most critical since the energy spectrum for
the $W$-daughter muons peaks at a value just
below the beam energy, falling off nearly linearly with decreasing
energy. One might also expect the hadron calorimeter energy
resolution to be important in that it affects the energy
resolution of jets to be identified with underlying $W$-daughter
quarks. Preliminary work indicates, however, that 
an NLC detector can tolerate a broad range in charged track
momentum and hadron calorimeter energy resolutions
without significant degradation of precision on extracted
anomalous couplings. This insensitivity to detector resolutions 
stems from the power of an over-constrained kinematic fit
in determining the five event angles.

One expects some degradation 
in coupling parameters precision from the ideal
case due to the underlying physical phenomena of initial state
radiation\cite{WARD} and the finite $W$ width and a smaller degradation
from the imperfection of matching detected particles to primary
$W$ daughters. The potentially largest effect 
comes from initial state radiation. 
With precise luminosity monitors, such as those 
in present LEP detectors,
for which luminosity \vs\ effective c.m. energy can be tracked,
straightforward corrections for ISR, including beamstrahlung,
should be feasible. One doesn't expect dramatic
degradation in sensitivity from any of the above complications.

Preliminary work suggests that the 4-jet channel
\jjjj\ can contribute significantly to anomalous
couplings measurements, as long as charge confusion in
the detector is well understood. In regard to ISR, this
channel has the important advantage of allowing 
reliable event-by-event
determination of missing photon energy, using kinematic
constraints, a technique that works only poorly in
the \jjlv\ channel\cite{KR}.
If one could reliably tag a $c(\bar c)$ quark jet, at least
one of which occurs in 75\% of the $\jjjj$-channel events,
one would expect further improvements in sensitivity.
This channel merits additional study. The purely leptonic \lvlv\ 
channel is remarkably clean, but poor statistics and 
ambiguous kinematics make this channel intrinsically less
sensitive than the \jjlv\ and \jjjj\ 
channels\cite{KALYNIAK}.

An NLC \epem\ collider also allows measurement of non-Abelian
gauge boson couplings in channels\cite{AGBI} other than
\epemww. 
The process $e^+e^-\rightarrow Z\gamma$
probes $ZZ\gamma$ and $Z\gamma\gamma$ couplings, and processes such as
$e^+e^- \rightarrow WWZ$ and $e^+e^- \rightarrow ZZZ$
probe quartic couplings\cite{GODFREY4,DAWSON4}.
The $WW\gamma$ and $WWZ$
couplings can be probed independently via the processes 
$e^+e^- \rightarrow \nu_e\bar\nu_e \gamma$ and 
$e^+e^- \rightarrow \nu_e\bar\nu_e Z$, 
respectively\cite{GODFREYGAMMA,GODFREYZ}.

Similar measurements can be carried out at \emem, \emg\ and \gg\
colliders, where the expected reduction in  
luminosity is at least partly compensated
by other advantages\cite{TLBFINLAND,CUYPERS}.
For example, the processes $\gamma e^-\rightarrow
W^-\nu_e$ and  $\gamma\gamma \rightarrow W^+W^-$ 
probe the $WW\gamma$ coupling, 
independent of $WWZ$ effects. The polarization asymmetry
in the former reaction reverses as the energy of the
collisions is varied, and the location of the zero-crossing
provides a probe of $\lamg$\cite{RIZZO}.
Similarly, the process $e^-\gamma\rightarrow Ze^-$ probes the 
$ZZ\gamma$ vertex\cite{RIZZO}.
In the process $e^-\gamma\rightarrow W^-Z\nu_e$, one
obtains sensitivity to $WWZ$, $WW\gamma$, and $WWZ\gamma$ couplings.
In particular, the parity-violating coupling $\gfivev$ can be
probed\cite{DAWSONPARITY}.
More power comes from the
ability to polarize {\it both} 
incoming beams with these alternative
colliders. An $e^-e^-$ collider has moreover 
the special capability of producing
isospin-2 intermediate states, 
such as in $e^-e^-\rightarrow\nu\nu W^-W^-$\cite{BARGER}.
The similar reactions $e^-e^-\rightarrow e^-\nu W^-Z$ and
$e^-e^-\rightarrow e^-e^- ZZ$ turn out to be powerful probes
for anomalous quartic couplings\cite{CUYPERS}.
Table~\ref{TABALT} shows a sampling of processes and
gauge couplings that can be
studied at alternate colliders associated with the NLC.

\begin{table*}[h]
\begin{center}
\caption{A sampling of processes and associated gauge boson couplings
measurable at $e^-e^-$, $\gamma\gamma$, and $e^-\gamma$
colliders.}
\label{TABALT}
\begin{tabular}{|l|l|}\hline\hline
Process & Couplings probed \\ \hline
$\eemr e^-\nu W^-$     & $WW\gamma$, $WWZ$            \\
$\eemr e^-e^- Z$       & $ZZ\gamma$, $Z\gamma\gamma$  \\
$\eemr e^-\nu W^-\gamma$ & $WW\gamma$, $WWZ$            \\
$\eemr \nu\nu W^-W^-$    & $WWWW$ \\
$\eemr e^-\nu W^-Z$    & $WWZZ$                   \\
$\eemr e^-e^- ZZ$    & $ZZZZ$  \\
\hline\hline
$\ggr W^+W^-$                & $WW\gamma$       \\
$\ggr W^+W^-Z$               & $WWZ$, $WW\gamma$ \\
$\ggr ZZ$                    & $ZZ\gamma$, $Z\gamma\gamma$ \\
$\ggr W^+W^-W^+W^-$          & $WWWW$  \\
$\ggr W^+W^-ZZ$              & $WWZZ$ \\
\hline\hline
$\egr W^-\nu$                & $WW\gamma$ \\
$\egr e^-Z$                  & $ZZ\gamma$, $Z\gamma\gamma$ \\
$\egr W^+W^-e^-$             & $WWZ$, $WW\gamma$, $WWZ\gamma$ \\ 
$\egr W^-Z\nu$             & $WWZ$, $WW\gamma$, $WWZ\gamma$ \\ \hline
\end{tabular}
\end{center}
\end{table*}

\section{Conclusions}

\begin{figure*}[b]
\leavevmode
\begin{center}
\includegraphics{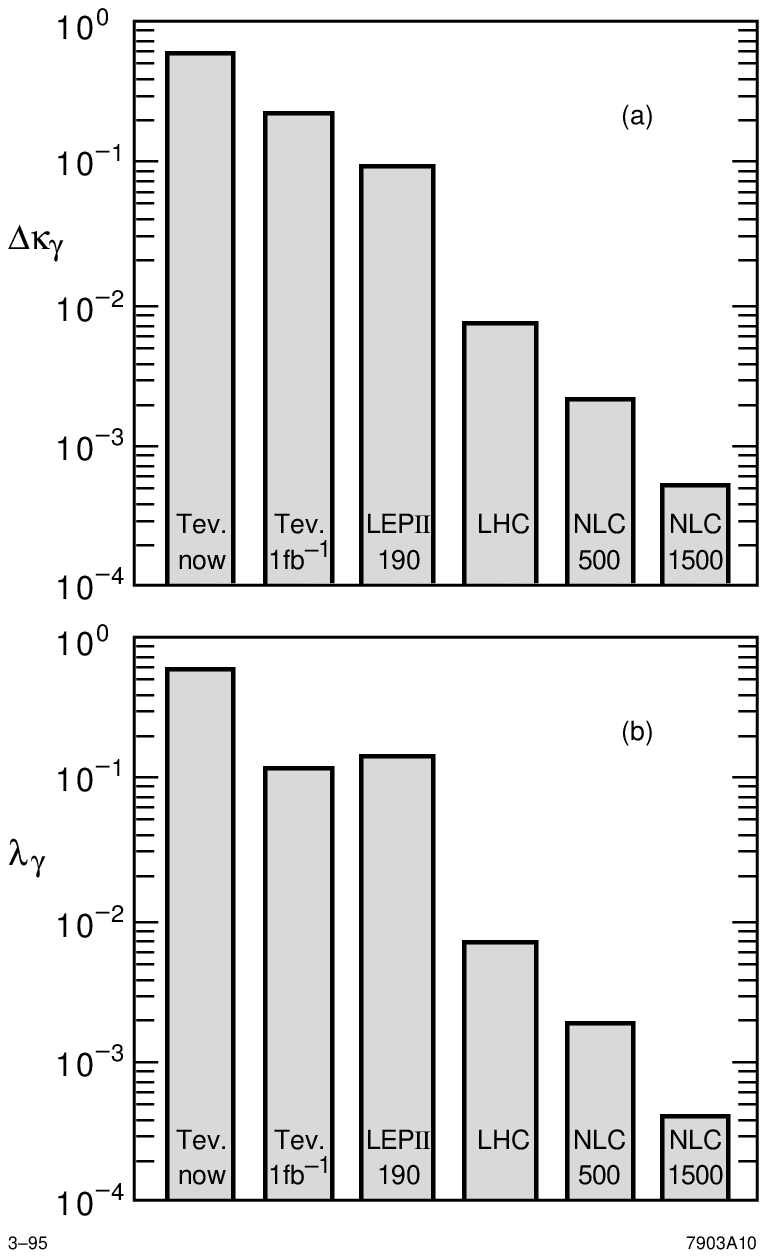}
\end{center}
\caption{Comparison of representative 95\% C.L. upper limits on
$\dkapg$ and $\lamg$ for present and future accelerators.}
\label{WWVFIGTHREE}
\end{figure*}

In the coming years, data from LEP II and an upgraded Tevatron
should provide sensitivities to various 
anomalous gauge boson couplings
of $O(10^{-1})$, an order of magnitude better than present
direct measurements from the Tevatron. 
The LHC should greatly
improve on this sensitivity, probing to better than $O(10^{-2})$.
An NLC at 500 GeV c.m. energy would do still better, reaching
$O(10^{-3})$, while a 1.5 TeV NLC would achieve sensitivities
of $O(10^{-4})$. Fig.~\ref{WWVFIGTHREE} 
(taken from ref.~\cite{BDHS}) shows a useful comparison among
these accelerators. The enormous potential of LHC and
especially that of a high-energy NLC are apparent. In general,
the LHC and NLC are complementary: the $e^+e^-$ collider (and
associated alternative $e^-e^-$, $e^-\gamma$, $\gamma\gamma$
colliders) allow precision measurement of helicity amplitudes
at well-determined c.m. energy, while the $pp$ collider allows
less precise probing of couplings at higher energies.
Both machines should probe energy scales of a few TeV and
should add decisively to
our understanding of gauge boson self interactions.

\end{document}